\documentstyle[aps,prl,floats,epsfig]{revtex}
\begin{document}

\title{\bf\large
Direct measurement of the $\phi(1020)$ leptonic 
branching ratio}

\author{
M.N.Achasov, 
K.I.Beloborodov,
A.V.Berdyugin, 
A.G.Bogdanchikov,
A.V.Bozhenok, 
A.D.Bukin, 
D.A.Bukin, 
S.V.Burdin\cite{byline},
T.V.Dimova, 
A.A.Drozdetsky, 
V.P.Druzhinin, 
M.S.Dubrovin, 
I.A.Gaponenko, 
V.B.Golubev, 
V.N.Ivanchenko, 
P.M.Ivanov,  
A.A.Korol, 
S.V.Koshuba, 
A.P.Lysenko,
I.N.Nesterenko,
E.V.Pakhtusova,
E.A.Perevedentsev, 
A.A.Salnikov, 
S.I.Serednyakov, 
V.V.Shary, 
Yu.M.Shatunov, 
V.A.Sidorov, 
Z.K.Silagadze, 
A.N.Skrinsky,
Yu.V.Usov, 
A.V.Vasiljev}
\address{Budker Institute of Nuclear Physics and
  Novosibirsk State University \\ 
 630090, Novosibirsk,  Russia }

\maketitle
\begin{abstract}
The process  $e^+e^-\to\mu^+\mu^-$ has been studied by SND detector
at VEPP-2M $e^+e^-$ collider in the  $\phi(1020)$-resonance energy region.
The measured effective $\phi$ meson leptonic branching ratio:
$B(\phi\to l^+l^-)\equiv\sqrt{B(\phi\to e^+e^-)\cdot B(\phi\to \mu^+\mu^-)}=(2.89\pm 0.10\pm 0.06)\cdot 10^{-4}$
agrees well with the PDG value
$B(\phi\to e^+e^-)=(2.91\pm 0.07)\cdot 10^{-4}$
confirming $\mu$--$e$ universality.
Without additional assumption of $\mu$--$e$ universality the
branching ratio 
$B(\phi\to \mu^+\mu^-)=(2.87\pm 0.20\pm 0.14)\cdot 10^{-4}$ 
was obtained.
\end{abstract}


Truly neutral vector mesons play an important role in hadron phy\-sics
due to their direct coupling to photons. This phenomenon is the basis of
the phenomenological Vector Meson Dominance model which successfully
describes electromagnetic interactions of hadrons.
The key parameters of this model are $V$--$\,\gamma$ coupling constants.
They can be extracted from the vector meson leptonic 
widths under the assumption that leptonic decay proceeds via
one-photon annihilation of the quark-antiquark pair constituting the 
meson.  
Leptonic widths also determine
the total production cross sections of vector mesons in $e^+e^-$
annihilation and are important for calculation of the hadronic contribution
to the photon vacuum polarization \cite{Eidelman}.

The $V$--$\,\gamma$ coupling constant is just one number
per vector meson. Could
these numbers tell us something non-trivial about the underlying QCD dynamics?
Shortly after the 1974 ``charm revolution'', Yennie noticed that
independently of the
vector meson flavor content the following relation holds
\cite{Yennie,Sakurai}
\begin{equation}
\Gamma(V\to e^+e^-)/<e_q>^2\approx 12~\mathrm{keV},
\label{Yenn} \end{equation}
\noindent where $<e_q>$ is the mean electric charge of the valence quarks
inside the vector meson $V$ in the units of an electron charge.
For $\rho,\;\omega$ and $\phi$ mesons this gives the famous 
rule: 
$\Gamma(\rho\to e^+e^-):\Gamma(\omega\to e^+e^-):\Gamma(\phi\to e^+e^-)=9:1:2,$
which can be considered as an SU(3) symmetry prediction.
The surprising fact 
here is a relatively high ($\sim 10\%$) precision 
of the 9:1:2 rule despite SU(3)-flavor symmetry
breaking. Inclusion of charm gives even more badly broken SU(4)
symmetry, but Yennie's relation remains valid with the same precision,
which means that SU(4) symmetry still persists for the leptonic widths ratios!
Inspired by this strange fact,
Gounaris predicted $\Gamma(\Upsilon\to e^+e^-)=1.2\mbox{keV}$ 
\cite{Gounaris} and was closer to reality than any other author
 \cite{Sakurai}. Current experimental situation with leptonic widths \cite{PDG}
 is shown in Table~\ref{tabvec}.

\narrowtext
\begin{table}[ht]
\caption{\protect The leptonic widths of vector mesons.
\label{tabvec}}
\begin{tabular}[t]{cccc}
&$\Gamma_{exp}$, keV&$<e_q>^2$&$\frac{\Gamma_{exp}}{<e_q>^2}$, keV \\
\tableline
$\rho$& $6.77\pm0.32$& $1/2$& $13.5\pm0.6$ \\
$\omega$& $0.60\pm0.02$& $1/18$& $10.8\pm0.4$\\
$\phi$&  $1.30\pm0.03$& $1/9$& $11.7\pm0.3$\\
$J/\psi$& $5.26\pm0.37$& $4/9$& $11.8\pm0.8$\\
$\Upsilon$& $1.32\pm0.05$& $1/9$& $11.9\pm0.5$ \\
\end{tabular}
\end{table}

\widetext
In the nonrelativistic potential model \cite{Quigg} the leptonic decay 
width is given by the Van Royen--Weisskopf formula \cite{Weiskopf}:
$\Gamma(V\to e^+e^-)=16\pi\alpha^2 <e_q>^2~|\Psi(r=0)|^2/M_V^2.$

Equation (\ref{Yenn}) implies then that quarkonium wave function at 
the origin $\Psi(r=0)$ is proportional to the meson mass $M_V$.
Note that for Coulomb potential $|\Psi(r=0)|^2\sim M_V^3$, while
the linear potential gives      $|\Psi(r=0)|^2\sim M_V$.
So the leptonic widths tell us
that the actual potential appears to be something in between.
But even if we postulate such a potential,
the relation (\ref{Yenn}) still has no
simple explanation. For light quark systems like $\rho$,
$\omega$, and $\phi$
relativistic corrections are essential. There are also strong interaction
corrections governed by the scale dependent $\alpha_s$. It was argued
\cite{Poggio} that these corrections modify the Van Royen - Weisskopf formula
in the following way:
\begin{equation}
\Gamma(V\to e^+e^-)\approx
16\pi\alpha^2 <e_q>^2~|\Psi(r=1/m_q)|^2 (1-0.36~\alpha_s(M_V))/M_V^2.
\label{VRWM} \end{equation} 
Intuitively, appearance of the constituent quark Compton wavelength
$1/m_q$ in (\ref{VRWM}) looks natural, because in relativistic theory a
particle cannot be localized within a region smaller than its
Compton wavelength \cite{Wigner}. Thus we can
expect the quark-antiquark pair to annihilate when approaching
each other's relativistic extents \cite{Poggio}. But
this intuitive clarity of (\ref{VRWM}) doesn't make an explanation of the
remarkable regularity of (\ref{Yenn}) simpler, because (\ref{VRWM}) shows that
leptonic widths are sensitive to the both nonperturbative and perturbative
aspects 
of QCD. Thus it is not surprising that the leptonic widths become
a traditional touchstone for various quark models \cite{Quigg,Quark}.

  This paper is devoted to the measurement of the leptonic branching ratio of
the $\phi(1020)$ meson. There are two leptonic decays:
$\phi\to e^+e^-$ and $\phi\to\mu^+\mu^-$. The $\mu$--$e$ universality implies 
for these decays that 
$B(\phi\to\mu^+\mu^-)=B(\phi\to e^+e^-)\times 0.9993$. 
Presently only the $\phi\to\mu^+\mu^-$ decay was measured directly 
\mbox{(\cite{Earles}--\cite{mu96})}. 
There are two PDG values for this decay branching ratio \cite{PDG}. 
One of them $B(\phi\to\mu^+\mu^-)=(2.5\pm 0.4)\cdot 10^{-4}$
is based on the experiments on photoproduction of $\phi$ meson 
\cite{Earles,Hayes}. Another value of the branching ratio 
$B(\phi\to\mu^+\mu^-)=(3.7\pm 0.5)\cdot 10^{-4}$
is obtained from $e^+e^-$ experiments \cite{Orsay,Chil,mu96}. 
In addition the CMD-2 experiment \cite{prep99-11,cmdhadr} 
has some preliminary results on this decay. 
One can see that the difference between
two PDG values for the decay $\phi\to\mu^+\mu^-$  is about 2 standard
deviations and the accuracy of these results is relatively low.
Current branching ratio $B(\phi\to e^+e^-)=(2.91\pm 0.07)\cdot
10^{-4}$ \cite{PDG} 
is based on measurements of the $\phi$-meson
total production cross section in $e^+e^-$ collisions. It was
obtained by summation of all $\phi$-meson decay modes: $\phi\to K^+K^-$,
$K_S K_L$, $3\pi$, etc..
Up to now the accuracy of $B(\phi\to e^+e^-)$ was much higher
than that of $B(\phi\to\mu^+\mu^-)$.
But there is a serious factor limiting the precision of
$B(\phi\to e^+e^-)$ obtained in such an indirect way.
It is the interference between $\phi$ meson and 
other vector states, which description is model dependent.
Direct measurement
of the $\phi\to e^+e^-$ decay in the $e^+e^-\to\phi\to e^+ e^-$ reaction
is difficult due to its small probability and huge background
from the $e^+e^-\to e^+e^-$ Bhabha scattering.

The decay $\phi\to \mu^+\mu^-$ reveals itself as a 
wave-like interference pattern
in the energy dependence of the $e^+e^-\to\mu^+\mu^-$ cross section
in the region close to the $\phi$-meson peak.
The amplitude of the interference wave is proportional to 
$B(\phi\to l^+l^-)\equiv\sqrt{B(\phi\to e^+e^-)\cdot B(\phi\to \mu^+\mu^-)}.$
The accuracy of the $B(\phi\to l^+l^-)$ measurement in this case is
limited only by uncertainty in the calculation of the pure QED part of the 
$e^+e^-\to\mu^+\mu^-$ cross section.
The 0.2\% accuracy claimed in \cite{theor} leads to 0.8\%
systematic error in the interference amplitude.
Large statistics collected by SND detector in the vicinity of
the $\phi$ resonance allowed us to make direct measurement of the
leptonic branching ratio $B(\phi\to l^+l^-)$ with the accuracy comparable
with that of previous indirect measurements of $B(\phi\to e^+e^-)$.

Our previous study of the $e^+e^-\to\mu^+\mu^-$ cross section was done
using the 1996 data sample  
with the total integrated luminosity  of $2.6$~pb$^{-1}$  \cite{mu96}.
In 1998 two experimental runs were carried out in the center of mass
energy range
$E=984-1060$~MeV in 16 energy points. The collider 
operated with superconducting wiggler \cite{snake} allowing to 
increase the average luminosity by a factor of two.
Higher luminosity led to 
relative reduction of the cosmic ray background.
The total integrated luminosity $\Delta L=8.6$~pb$^{-1}$ collected in 1998 
corresponds to $13.2\cdot 10^{6}$ produced $\phi$ mesons. 

  The SND experimental setup is described in detail in 
ref.~\cite{SND}.
The main part of the SND is a spherical electromagnetic calorimeter.
The angles of charged particles  are measured by two cylindrical drift
chambers (DC). An outer
muon system, consisting of streamer tubes and plastic scintillation counters,
covers the detector.
The integrated
luminosity was measured using $e^+e^-\to e^+e^-$ events selected in
the same acceptance angle as the events of the process under study
$e^+e^-\to\mu^+\mu^-$.
The systematic uncertainty of the luminosity measurement is 2\%, 
but its contribution to the systematic error of the 
interference amplitude estimated using the process 
$e^+e^-\to \gamma\gamma$ is only 0.8\%. 

The primary selection criteria for $\mu^+\mu^-$
events were similar to those of our previous work \cite{mu96}:
\begin{itemize}
\item total energy deposition in the calorimeter is more than 270~MeV;
\item there are two collinear charged tracks in an event with
acollinearity angles in azimuth and polar directions:
 $\mid\Delta\varphi\mid<10^\circ$, $\mid\Delta\theta\mid<25^\circ$
and with the polar angles within
$45^\circ<\theta<135^\circ$ ;
\item event is not tagged as $e^+e^-\to e^+e^-$ by $e/\pi$
 separation procedure \cite{pi98}.
\end{itemize}
  To suppress the background from the processes
  $e^+e^-\to\pi^+\pi^-$, $\pi^+\pi^-\pi^0$, $K_SK_L$, $K^+K^-$ the outer
  muon system was used: a requirement for both charged particles 
to produce \mbox{hits} in muon system renders contribution from this
background negligible. For example, the contribution from the process 
$e^+e^-\to\pi^+\pi^-$ is about 0.2\% in the $\phi$-meson peak.

\begin{figure}
\begin{minipage}{0.48\textwidth}
\epsfig{figure=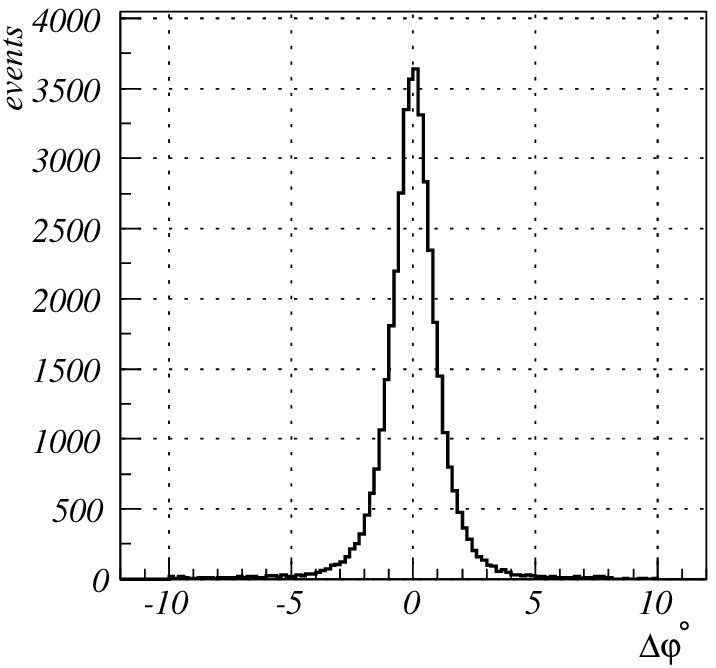,width=0.99\textwidth}
\caption{
The $\Delta\varphi$ distribution in $e^+e^-\to\mu^+\mu^-$ events.}
\label{dphm}
\end{minipage}
\hfill
\begin{minipage}{0.48\textwidth} 
\epsfig{figure=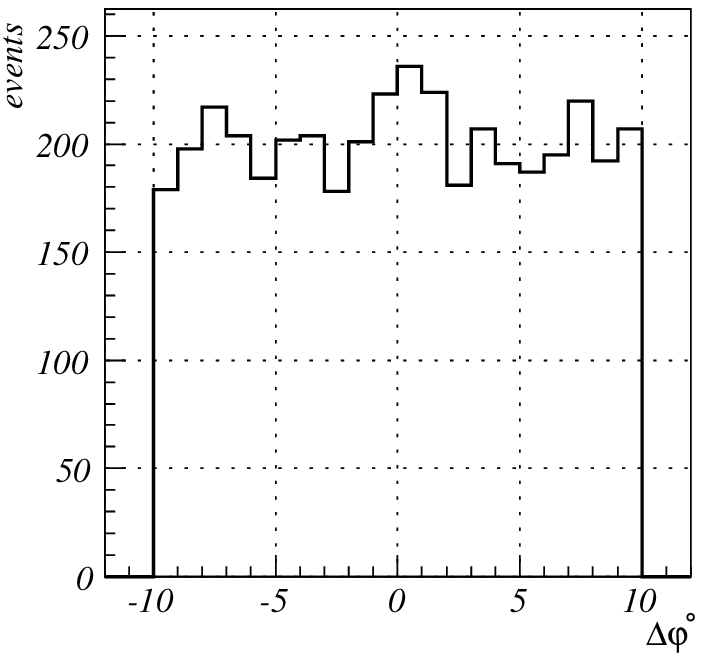,width=0.99\textwidth}
\caption{
The $\Delta\varphi$ distribution for cosmic ray events. }
\label{dphc}
\end{minipage}
\end{figure}

  The cosmic ray background was suppressed by restriction of the
  time $\tau$ measured by outer scintillation counters
  with respect to the beam collision moment \cite{mu96}:
  $\mid\tau\mid<10$~ns. 
  About 30\% of events selected by the \mbox{cuts} described
above are still cosmic ray background.
To determine the contribution of cosmic background more accurately
the selected events were divided into two classes: (1) 
$\mid\Delta\varphi\mid<5^\circ$; (2) $\mid\Delta\varphi\mid>5^\circ$.
The resolution in $\Delta\varphi$ is about $1^\circ$.
The  $\Delta\varphi$ distribution for
$e^+e^-\to\mu^+\mu^-$ events (Fig.~\ref{dphm})
was obtained from the
experimental data after strong \mbox{cuts} on a difference between time
measurements by the muon system for both tracks.
Almost all $\mu^+\mu^-$ events belong to the first class.
The second class contains only
$1.7\%$ of $\mu^+\mu^-$ events.
The $\Delta\varphi$ distribution for pure cosmic ray events collected 
in a special run without beams is shown in Fig.~\ref{dphc}.
The uniformity of this distribution is an artifact of our DC track
reconstruction algorithm in which the origin of a charged track in the
$X$-$Y$ plane is fixed to the beam collision point.
From Fig.~\ref{dphc} the ratio between numbers of cosmic ray events in the two 
classes was found
$k_{cs}=N^{cs}_1/N^{cs}_2=1.028\pm0.033.$

The number of cosmic ray background events in the first class
was calculated
for each energy point $E_i$ by the following formula:
$ N^{cs}_1(E_i)=k_{cs}\cdot T(E_i)\cdot\mathrm{d}N_2/\mathrm{d}T. $
Here $T(E_i)$ is a data acquisition time for an energy
point $E_i$, $\mathrm{d}N_2/\mathrm{d}T$ is
the cosmic event rate in class two 
averaged over  both 1998 experimental runs.
The net number of $\mu^+\mu^-$ events for each energy point
was obtained by subtraction of the cosmic ray background:
$ N^\mu (E_i)=N_1(E_i)-N^{cs}_1(E_i).$ The errors of the numbers 
$N^\mu (E_i)$ include the errors of $N_1(E_i)$ and $N^{cs}_1(E_i)$.

Energy dependence of the detection cross section was fitted
according to the following formula:
\begin{eqnarray}
 \sigma^{vis}_{\mu\mu}(E) & = &\sigma_0(E)\cdot R(E)\left| 1 -
 Z_\mu\frac{m_{\phi}\Gamma_{\phi}}{\Delta_{\phi}(E)}\right|^2, 
 \nonumber \\
\label{crossmu} 
 \sigma_0(E) & = & 2\pi\alpha^2\beta(E)(1-\beta^2(E)/3)/E^2, 
\end{eqnarray}
where $\alpha$ is the fine structure constant;
$\beta(E)=(1-4\cdot m^2_\mu/E^2)^{1/2}$;
$m_\phi$, $\Gamma_{\phi}$,
$\Delta_{\phi}(E)
= m_{\phi}^{2} - E^2 - iE\Gamma(E)$ are the
$\phi$-meson mass, width and inverse  propagator respectively; 
$\sigma_0(E)$ is Born
cross section of the process $e^+e^-\to\mu^+\mu^-$; 
$Z_\mu \equiv Q_\mu e^{i\psi_\mu}$ --- interference parameter.
The modulus of the interference parameter is related to the leptonic 
branching ratio:
$Q_\mu=B(\phi\to l^+l^-)\cdot 3/\alpha.$
The factor $R(E)$ takes into account
the detection efficiency and radiative corrections:
\begin{equation}
R(E)=\varepsilon_\mu\frac{\sigma_{\mu\mu}(E)}{\sigma_0(E)
\left|1-Z\frac{m_{\phi}\Gamma_{\phi}}{\Delta_{\phi}(E)} \right| ^2}.
\label{Rfac}
\end{equation}
Here $\sigma_{\mu\mu}$ is the result of Monte Carlo integration of the
differential cross section of the process
$e^+e^-\to\mu^+\mu^-(\gamma)$ \cite{theor} for our geometric \mbox{cuts} with
the energy dependent probability for muons to hit
outer scintillation counters taken into account. 
The uncertainty in the energy dependence of this probability 
adds 1.7\% to the systematic error of $Q_\mu$. 
The parameter $\varepsilon_\mu$ represents energy independent 
factor in the detection efficiency. It is determined mainly 
by the cut on total energy deposition in the  calorimeter.
The value of $\varepsilon_\mu=0.84\pm 0.01$ was obtained using Monte Carlo
simulation of the process $e^+e^-\to\mu^+\mu^-(\gamma)$ in SND
detector \cite{UNIMOD}, but in the fitting procedure $\varepsilon_\mu$
was left free. In the calculation of
radiative corrections the interference parameter was assumed
purely real and equal to
$Z=B(\phi\to e^+e^-)\cdot 3/\alpha=0.120$
with the PDG table value for $B(\phi\to e^+e^-)$. 
\begin{figure}
\begin{minipage}[htb]{0.99\textwidth}
\epsfig{figure=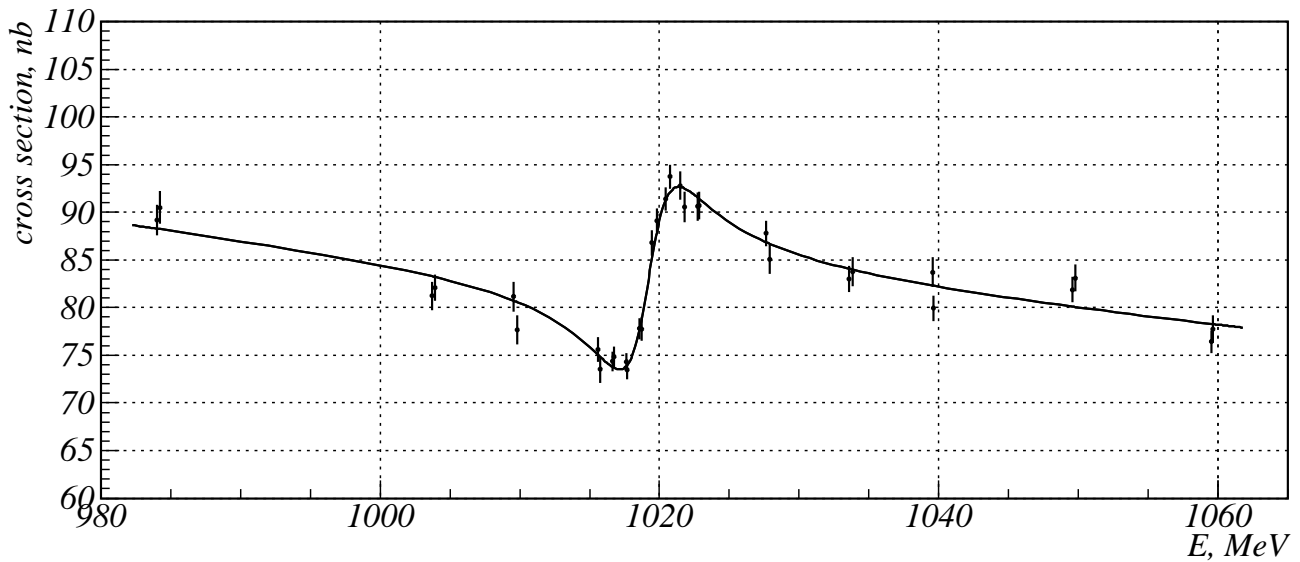,width=0.99\textwidth}
\caption{
The measured cross section of the process
$e^+e^-\to\mu^+\mu^-$.
}
\label{secall}
\end{minipage}
\end{figure}

\begin{table}
\caption
{
The results of the fit with $\psi_\mu=0$ for two experimental runs. Only
statistical errors are shown.
}
\label{tabres}
\begin{tabular}{cccc}
Parameter & PHI\_9801   &  PHI\_9802  & Combined \\
\tableline
$\chi^2/NDF$ & $19.4/15$ &  $ 11.3/15$   &  $33.8/30$      \\
$Q_\mu,\; 10^{-2}$ & $12.1 \pm 0.6$ & $11.0\pm 0.6$ & $11.9\pm 0.4$\\
$\varepsilon_\mu,\; 10^{-2}$ & $ 83.1\pm0.3$ & $ 82.5\pm0.3$ &
$83.1(82.8)\pm0.3$ \\
$B(\phi\to l^+l^-),\; 10^{-4}$& $ 2.99\pm0.15$
& $ 2.74\pm0.14$ & $ 2.89\pm0.10$ \\
\end{tabular}
\end{table}

The fitting was performed for two experimental runs independently.
Fits with a free $\psi_\mu$ yield the interference phase, which is 
in good agreement with the expected zero value:
(1) $\psi_\mu=(1.0\pm 2.8)^\circ$, (2) $\psi_\mu=(0.1\pm 2.8)^\circ$.
Therefore the interference phase was fixed to
$\psi_\mu=0$. The fit results presented in Table~\ref{tabres} show
statistical agreement between two experimental runs.  
Therefore combined fit was performed to obtain the final results which are
listed in the third column of the Table~\ref{tabres}. 
The values of $\varepsilon_\mu$ obtained in the fit and from Monte
Carlo simulation agree well. The energy dependence 
of the measured cross section and the fitting curve are shown in 
Fig.~\ref{secall}. The systematic error of $Q_\mu$ includes 1.7\% from
the uncertainty in the energy dependence of the probability for muons
to hit the outer system, 0.8\% from the luminosity measurements and
0.8\% from the calculation of the radiative corrections. The resulting
systematic error is 2\%.

In conclusion we obtain the following $\phi$ meson parameters 
from the measured $Q_\mu$ value:
$B(\phi\to l^+l^-)=(2.89\pm 0.10\pm 0.06)\cdot 10^{-4}$;
$B(\phi\to e^+e^-)\cdot B(\phi\to\mu^+\mu^-)=(8.36\pm0.59\pm0.37)\cdot 10^{-8}.$
This result is in good agreement with our previous one $B(\phi\to
l^+l^-)=(3.14\pm 0.22\pm 0.14)\cdot 10^{-4}$ \cite{mu96}. 
Using PDG value of $B(\phi\to e^+e^-)=(2.91\pm0.07)\cdot 10^{-4}$ we
obtain: 
$B(\phi\to\mu^+\mu^-)=(2.87\pm0.20\pm 0.14)\cdot 10^{-4}.$
The good agreement of  
$B(\phi\to\mu^+\mu^-)$ and $B(\phi\to e^+e^-)$ confirms the 
$\mu$--$e$ universality. 

This work is supported in part by Russian Foundation for Basic 
Research, grants 
No. 99-02-16815 and 99-02-17155.
We thank G.V.Fedotovich for useful discussions.


\begin {references}
\bibitem[*]{byline} E-mail: burdin@inp.nsk.su
\bibitem{Eidelman}
S.~Eidelman, F.~Jegerlehner, Z.~Phys. C {\bf 67}, 585 (1995).
\bibitem{Yennie} D.~R.~Yennie, Phys. Rev. Lett. {\bf 34}, 239 (1975).
\bibitem{Sakurai} J.~J.~Sakurai, Physica {\bf A96}, 300 (1976). 
\bibitem{Gounaris} G.~J.~Gounaris, Phys. Lett. B {\bf 72}, 91 (1977). 
\bibitem{PDG}
Particle Data Group, D.~E.~Groom {\it et al.,} Euro. Phys. J. {\bf C15}, 
1 (2000).
\bibitem{Quigg} C.~Quigg, J.~L.~Rosner, Phys. Rept. {\bf 56}, 167 (1979);
H.~Grosse, A.~Martin, Phys. Rept. {\bf 60}, 341 (1980).
\bibitem{Weiskopf}  R.~Van Royen, V.~F.~Weisskopf,  Nuovo Cim. 
{\bf A50}, 617 (1967); {\it ibid.} {\bf A51}, 583 (1967).
\bibitem{Poggio} E.~C.~Poggio, H.~J.~Schnitzer, Phys. Rev. D {\bf 20},
1175 (1979); {\it ibid.} {\bf 21}, 2034 (1980);
H.~Krasemann, Phys. Lett. B {\bf 96}, 397 (1980).
\bibitem{Wigner} T.~D.~Newton, E.~P.~Wigner, Rev. Mod. Phys. {\bf 21}, 400 
(1949).
\bibitem{Quark} see, for example, W.~Jaus, Phys. Rev. D {\bf 44}, 2851 
(1991); B.~Margolis, R.~R.~Mendel, Phys. Rev. D {\bf 28}, 468 (1983);
N.~Barik, P.~C.~Dash, A.~R.~Panda, Phys. Rev. D {\bf 47}, 1001 (1993);
{\it ibid.} {\bf 53}, 4110 (1996);
P.~Maris, P.~C.~Tandy, Phys. Rev. C {\bf 60}, 055214 (1999);
B.~D.~Jones, R.~M.~Woloshyn, Phys. Rev. D {\bf 60}, 014502 (1999);
B.~C.~Metsch, H.~R.~Petry, Acta Phys. Polon. B {\bf 27}, 3307 (1996).
\bibitem{Earles}
 D.R.Earles {\it et al.,} Phys. Rev. Let. {\bf 25}, 1312 (1970). 
\bibitem{Hayes}
 S.Hayes {\it et al.,} Phys. Rev. D {\bf 4}, 899 (1971). 
\bibitem{Orsay}
 J.E.Augustin {\it et al.,} Phys. Rev. Lett. {\bf 30}, 462 (1973).
\bibitem{Chil}
 I.B.Vasserman {\it et al.,} Phys. Lett. B {\bf 99}, 62 (1981).
\bibitem{prep99-11}
 R.R.Akhmetshin {\it et al.,} Preprint BINP {\bf 99-11} (1999).
\bibitem{cmdhadr}
 R.R.Akhmetshin {\it et al.,}
 Nucl. Phys. A {\bf 675}, 424c (2000).
\bibitem{mu96}
 M.N.Achasov {\it et al.,} Phys. Let. B {\bf  456}, 304 (1999).
\bibitem{snake}
 V.V.Anashin {\it et al.,} Preprint BINP {\bf 84-123} (1984). 
\bibitem{SND} 
 M.N.Achasov {\it et al.,} Nucl. Inst. and Meth. A {\bf 449}, 125 (2000).
\bibitem{pi98}
 M.N.Achasov {\it et al.,} Phys. Let. B {\bf 474}, 188 (2000).
\bibitem{theor}
 A.B.Arbuzov {\it et al.,} Large angle QED processes
 at $e^+e^-$ colliders at energies below 3~GeV, hep-ph/9702262,
 JHEP 9710,  001 (1997).
\bibitem{UNIMOD}
 A.D.Bukin {\it et al.,} Preprint BINP {\bf 90-93} (1990).
\end {references}
\end{document}